\documentclass[proof]{WileyASNA-v1}

\articletype{IBWS proceedings}%

\received{30 April 2018}
\revised{***}
\accepted{***}

\begin{document}

\title{A stellar fly-by close to the Galactic centre: Can we detect stars on highly-relativistic orbits?}

\author[1,2]{Michal Zaja\v{c}ek*}

\author[3]{Arman Tursunov}

\authormark{M. Zaja\v{c}ek and A. Tursunov}

\address[1]{Max-Planck-Institut f\"ur Radioastronomie (MPIfR), Auf dem H\"ugel 69, D-53121 Bonn, Germany}

\address[2]{I. Physikalisches Institut der Universit\"at zu K\"oln, Z\"ulpicher Strasse 77, D-50937 K\"oln, Germany}

\address[3]{Institute of Physics and Research Centre of Theoretical Physics and Astrophysics, Faculty of Philosophy and Science,\\Silesian University in Opava, Bezru\v{c}ovo n\'{a}m.13, CZ-74601 Opava, Czech Republic}

\corres{*Corresponding author. \email{zajacek@ph1.uni-koeln.de}}

\abstract{The Galactic centre Nuclear Star Cluster is one of the densest stellar clusters in the Galaxy. The stars in its inner portions orbit the supermassive black hole associated with compact radio source Sgr~A* at the orbital speeds of several thousand km/s. The B-type star S2 is currently the best case to test the general relativity as well as other theories of gravity based on its stellar orbit. Yet its orbital period of $\sim 16\,{\rm yr}$ and the eccentricity of $\sim 0.88$ yields the relativistic pericentre shift of $\sim 11'$, which is observationally still difficult to reliably measure due to possible Newtonian perturbations as well as reference-frame uncertainties. A naive way to solve this problem is to find stars with smaller pericentre distances, $r_{\rm p}\lesssim 1529$ Schwarzschild radii ($120\,{\rm AU}$), and thus more prominent relativistic effects. In this contribution, we show that to detect stars on relativistic orbits is progressively less likely given the volume shrinkage and the expected stellar density distributions. Finally, one arrives to a sparse region, where the total number of bright stars is expected to fall below one. One can, however, still potentially detect stars crossing this region. In this contribution, we provide a simple formula for the detection probability of a star crossing a sparse region. We also examine an  approximate time-scale on which the star reappears in the sparse region, i.e. a ``waiting'' time-scale for observers.}


\jnlcitation{\cname{%
\author{M. Zaja\v{c}ek}, and 
\author{A. Tursunov}, 
} (\cyear{2018}), 
\ctitle{A stellar fly-by close to the Galactic centre: Can we detect stars on highly-relativistic orbits?}, \cjournal{Astronomical Notes}, \cvol{2018;00:1--6}.}

\fundingInfo{}

\maketitle


\section{Introduction}

The Galactic centre Nuclear Star Cluster (NSC) is considered as a laboratory for studying stellar dynamics in the dense stellar environment \citep{2000MNRAS.317..348G,2009A&A...502...91S,2013degn.book.....M,2014CQGra..31x4007S}. It is the only galactic nucleus, in which we can study the proper motion and radial velocities of individual stars inside the gravitational influence radius of Sgr~A* associated with the supermassive black hole (SMBH),

\begin{align}
  r_{\rm h} &= GM_{\bullet}/\sigma_{\star}^2\,,\notag\\
            &\approx 1.72\,\left(\frac{M_{\bullet}}{4\times 10^6\,M_{\odot}}\right)\left(\frac{\sigma_{\star}}{100\,{\rm km\,s^{-1}}} \right)^{-2} \,{\rm pc}\,,
\end{align}
which is derived by setting the characteristic circular velocity of stars bound to the black hole, $v_{\rm K}=\sqrt{GM_{\bullet}/r}$, equal to the one-dimensional, often line-of-sight, stellar velocity dispersion in the NSC, $\sigma_{\star}$. For $r\lesssim r_{\rm h}$, the gravitational potential of the SMBH prevails over the one of the NSC and the Keplerian rise in orbital velocities can be detected, $v_{\rm K} \propto r^{-1/2}$. At the Galactic centre distance of $\sim 8\,{\rm kpc}$, the angular scale of the influence radius is $\theta_{\rm h}\simeq 43''$, which can be resolved out to the angular scales of $\theta_{\rm min} \approx 63\,(\lambda/2\,{\rm \mu m})\,{\rm mas}$ for diffraction-limited observations in the NIR $K_{\rm s}$-band with eight-meter class telescopes \citep{2005bhcm.book.....E}. An even larger angular resolution is nowadays achieved with the Very Large Telescope Interferometer (VLTI), in particular GRAVITY@ESO instrument \citep{2011Msngr.143...16E}, which performs precision narrow-angle astrometry of the order of $10\,{\rm \mu as}$ as well as the phase-referenced interferometric imaging with the angular resolution of $4\,{\rm mas}$.  

\begin{figure*}[h!]
  \centering
  \includegraphics[width=\textwidth]{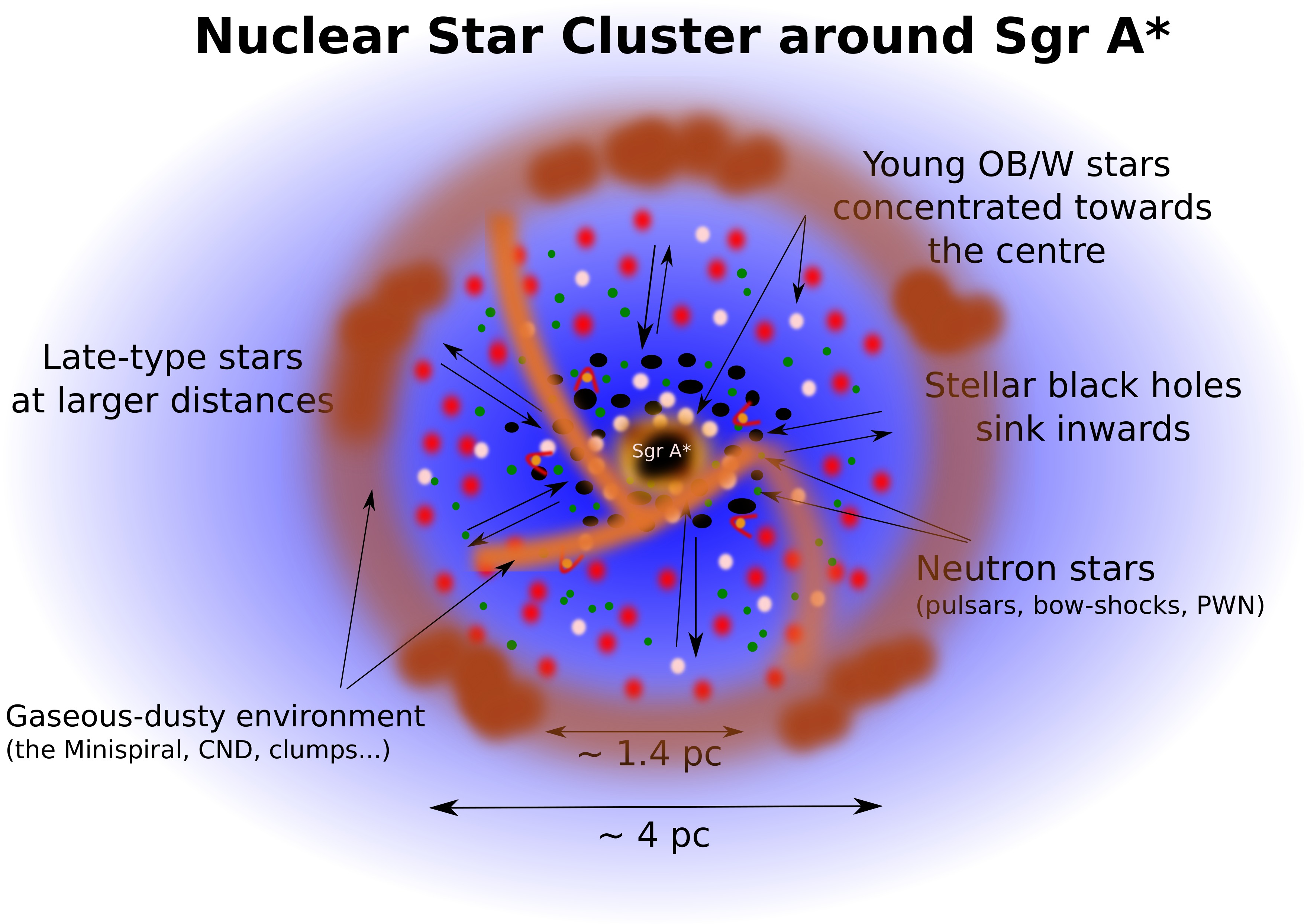}
  \caption{Illustration of different components of the NSC: old, late-type stars decrease towards the centre, forming a flatter core, while young OB stars seem to increase in numbers in the same direction, forming a cusp. In the central $\sim 2\,{\rm pc}$, prominent gaseous-dusty structures are located, mainly the neutral and molecular CND as well as ionized streamers of the Minispiral.}
  \label{img_nsc}
\end{figure*}

Based on the stellar counts in the central $\lesssim 2\,{\rm pc}$, the stellar mass and number density of late-type stars can be in general fitted by a broken power-law \citep{2009A&A...499..483B},

\begin{equation}
  \rho_{\star}=\rho_{0} \left(\frac{r}{r_{\rm b}} \right)^{-\gamma} \left[1+\left(\frac{r}{r_{\rm b}} \right)^{\delta} \right]^{(\gamma-\gamma_0)/\delta}\,,
  \label{eq_broken_power_law}
\end{equation}
where $\gamma$ is the slope of the inner distribution, $\gamma_0$ of the outer one, $\delta$ is the sharpness of the transition. The normalization constant $\rho_0$ is set according to the measured total enclosed mass of the NSC at a certain distance. \citet{2009A&A...502...91S} determined that the enclosed mass in the inner parsec is in the range $M_{\star}(<1\,{\rm pc})=0.5-1.0\times 10^6\,M_{\odot}$. The observed stellar distribution is consistent with the power-law model for $\rho_0=5.2 \times 10^{-5}\,M_{\odot}\,{\rm pc^{-3}}$, $r_0\approx 0.5\,{\rm pc}$, $\gamma=0.5$, $\gamma_0=1.8$, and $\delta=2$. This describes the fact that the late-type stars exhibit a flat core in the inner $r_{\rm b}\approx 0.5\,{\rm pc}$. The density of early-type stars, on the other hand, increases towards the compact radio source Sgr~A*, forming a cusp \citep{2009A&A...499..483B}. The illustration of the distribution is in Fig.~\ref{img_nsc}, which shows different stellar populations -- late-type stars with a core and early-type stars with a cusp as well as stellar remnants. Moreover, denser gaseous-dusty structures are located in the same region as the NSC and they orbit around the Galactic centre in a quasi-Keplerian way. The inner edge of the neutral and molecular Circum-Nuclear Disc (CND) coincides with the radius of the SMBH sphere of influence, $r_{\rm inner}^{\rm CND}\sim 1.7\,{\rm pc}$ \citep{1989MNRAS.240..219D}, inside which the ionized hot plasma is located that emits thermal X-ray bremsstrahlung \citep{2018MNRAS.474.3787M}. Three ionized denser streamers of Sgr~A~West or the Minispiral are thought to have dynamically originated in the CND via the loss of the angular momentum due to mutual collisions of CND clumps or the interaction with fast stellar winds of massive OB stars \citep{2000NewA....4..581V,2017A&A...603A..68M}.

Of a particular interest is the inner cluster within the projected radius of $\sim 1''\simeq 0.04\,{\rm pc}$ that consists of $\sim 30$ mostly B-type bright stars -- so-called S cluster \citep{1996Natur.383..415E,1997MNRAS.284..576E,1998ApJ...509..678G,2009ApJ...692.1075G,2017ApJ...837...30G,2017ApJ...847..120H}. These stars have nearly isotropic distribution of elliptical orbits with considerable eccentricities and pericentre distances of $\sim 1500$ Schwarzschild radii and orbital velocities at the periapse of several $1000\,{\rm km\,s^{-1}}$. One of the brightest stars S2 has the orbital period of $\sim 16$ years and it was possible to take measurements of its proper motions and radial velocities along its whole orbit. Thanks to the long-term monitoring of S2 and other two stars (S38 and S55/S0-102), it was possible to put the first weak constraints on the periastron shift of S2, which so far agrees with the relativistic prediction \citep{2017ApJ...845...22P}. It should be noted that the first-order post-Newtonian effects can also be revealed in the orbital radial velocities, when the near-infrared spectroscopic data is available. For the S2 star, the general relativistic radial-velocity shift per orbit is $<\Delta V^{\rm GE}> -11.6\,{\rm km\,s^{-1}}$ \citep{2017MNRAS.472.2249I}.      

Although the number density of stars is in general assumed to increase as $n_{\star}(r)\approx n_{0}(r/r_0)^{-\gamma}$, where $\gamma \geq 0$, the total number of stars inside radius $r$, $N_{\star}(<r)$, will decrease for $0\leq \gamma < 3$. The maximum value of $\gamma$ as inferred from the infrared observations typically reaches $\gamma_{\rm max}\approx 2$ \citep{2009A&A...502...91S}, hence the total number of stars within the sphere of radius $r$ must necessarily decrease, just because the total volume scales with $r^{3}$.

The number of stars inside the given radius $r$ can then be calculated as follows,

\begin{equation}
  N_{\star}(<r)=\int_0^r n(r') 4\pi r'^2 \mathrm{d}r=\frac{4\pi n_0}{r_0^{-\gamma}(3-\gamma)}r^{3-\gamma}\,.
  \label{eq_numberstars}
\end{equation}

For the number of stars inside the influence radius $N_{\star}(<r_{\rm h})=N_{\rm h}$, we obtain the analogical expression to Eq.~\eqref{eq_numberstars}. Hence, the general expression can be normalized with respect to the influence radius $r_{\rm h}$ in the following way,

\begin{equation}
   N_{\star}(<r) = N_{\rm h} \left(\frac{r}{r_{\rm h}} \right)^{(3-\gamma)}\,,
   \label{eq_numberstars_r}
\end{equation} 
which can be inverted to obtain the radius, within which there is a total number of $N_{\star}$ stars,

\begin{equation}
   r_{N_{\star}} = r_{\rm h} \left(\frac{N_{\rm h}}{N_{\star}(<r)} \right)^{-\frac{1}{3-\gamma}}\,.
   \label{eq_radius_N}
\end{equation} 

In this research note, we define the sparse region (hereafter SR) around Sgr~A* with radius $r_{\rm sparse}\lesssim r_{1}$, which is expected to contain less than one star, where $r_1$ directly follows from Eq.~\eqref{eq_radius_N}, by setting $N_{\star}(<r)=1$,

\begin{equation}
   r_{1} = r_{\rm h} N_{\rm h}^{-\frac{1}{3-\gamma}}\,.
   \label{eq_radius_1}
\end{equation} 

\begin{figure}[h!]
  \centering
  \includegraphics[width=0.5\textwidth]{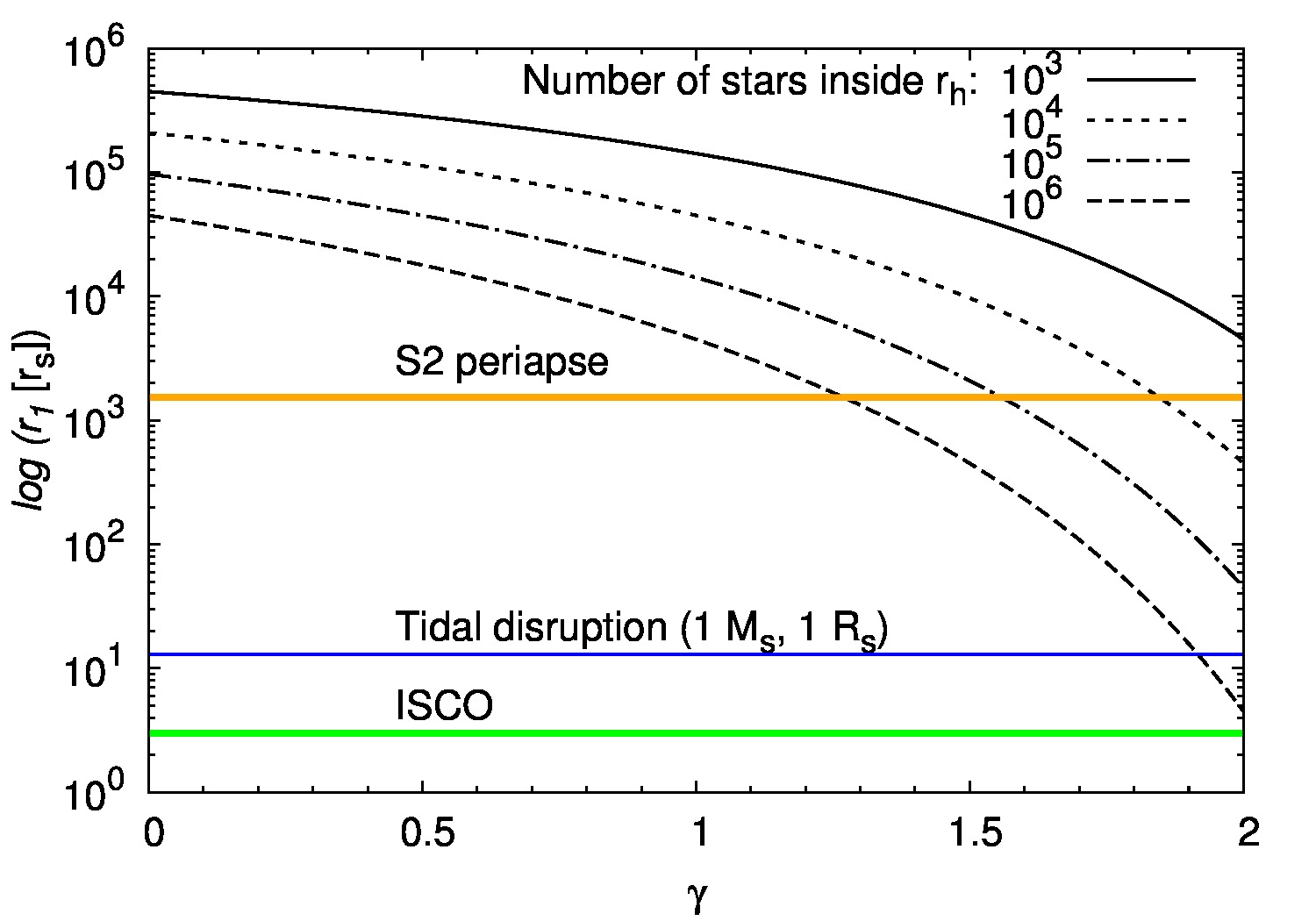}
  \caption{The radius of the sparse region $r_1$, see Eq.~\eqref{eq_radius_1}, inside which the number of stars is expected to be less or equal to one. The power-law slope of the stellar population is varied continuously in the range $\gamma = (0,2)$ and the number of objects inside the influence radius is increased in discrete steps by an order of magnitude, $N_{\rm h}=[10^3,10^4,10^5,10^6]$.}
  \label{img_radius_1}
\end{figure}

Equation~\eqref{eq_radius_1} implies that the radius of the SR depends strongly on the number of objects of particular type (late-type stars, early-type stars, compact objects) as well as their power-law slopes. In Fig.~\ref{img_radius_1}, we calculate $r_1$ for different numbers of objects inside $r_{\rm h}$, $N_{\rm h}=[10^3,10^4,10^5,10^6]$, and expected power-law slopes in the range $\gamma = (0,2)$.

It is useful to compare $r_1$ with characteristic length-scales in the inner parts of the NSC, where relativistic effects may become important. If we assume that Sgr~A* is a black hole, its Schwarzschild radius is (for a non-rotating case) $r_{\rm s}=2GM_{\bullet}/c^2\simeq 1.18 \times 10^{12} \left(M_{\bullet}/4\times 10^6\,M_{\odot} \right)\,{\rm cm}$. The tidal disruption of stars close to the black hole is expected at the tidal radius,

\begin{align}
   r_{\rm t} &= R_{\star}\left(\frac{3M_{\bullet}}{M_{\star}}\right)^{1/3}\,\notag\\
             &\approx 13 \left(\frac{R_{\star}}{R_{\odot}} \right) \left(\frac{M_{\star}}{M_{\odot}}\right)^{-1/3} r_{\rm s}\,.
\end{align}
The observationally important length-scale is the periapse distance of the brightest star S2. With the semimajor axis of $a=0.126''$ and the eccentricity of $e=0.884$ \citep{2017ApJ...845...22P}, the periapse distance is $r_{\rm p}=a(1-e)\approx 1529\,r_{\rm s} = 120.6\,{\rm AU}$.
These length-scales are plotted in Fig.~\ref{img_radius_1} alongside different profiles of $r_1$ as a function of the power-law slope $\gamma$. We see that only for steeper stellar density profiles, $\gamma>1$, the radius $r_1$ reaches the S2 periapse distance, which implies that inside S2 periapse distance the number of stars can be of the order of unity, as we will discuss in more detail in the following section.

\section{Analysis of a detection probability in a sparse region}

Even though the density of stars and stellar remnants in the Galactic centre is one of the largest in the Galaxy, the total number of stars falls below one at a certain distance from Sgr~A* due to the finite number of stars. \citet{2006ApJ...645.1152H} calculated the inner radius where the stellar cusp ends for different stellar components or types of objects (general notation $T$): main-sequence stars (MS), white dwarfs (WD), neutron stars (NS), and black holes (BH). Their general relation is merely an adjustment of our Equation~\eqref{eq_radius_1} to specific stellar types,

\begin{equation}
  r_{1,T}=(C_{T} N_{h})^{-1/(3-\gamma_{\rm{T}})} r_{\rm{h}}\,,
  \label{eq_inner_radius}
\end{equation} 
where $N_{\rm{h}}$ is the total number of MS stars, $C_{T} N_{h}$ is the total number of stars of type $T$ within the radius of influence of the black hole $r_{\rm{h}}$, and $\gamma_{T}$ is the power-law exponent for stellar type $T$. According to \citet{2006ApJ...645.1152H} the total number of MS stars within the radius of the gravitational influence $r_{h}=1.7\,\rm{pc}$ is $N_{h}=3.4\times 10^6$. Table~\ref{tab_inner_radius} summarizes the inner radii of different stellar populations of the NSC.

\begin{table}
\centering
  \begin{tabular}{ccccc}
  \hline
  \hline
  Population $T$ & $C_{T}$ & $\gamma_{T}$ & $r_{1,T}\,[\rm{pc}]$ & $r_{1,T}\,[r_{\rm s}]$ \\
  \hline
  MS                     & $1$       & $1.4$          & $2\times 10^{-4}$    &  $523$    \\  
  WD                     & $10^{-1}$    & $1.4$          &  $7\times 10^{-4}$ & $1831$       \\
  NS                     & $10^{-2}$   & $1.5$          &  $2\times 10^{-3}$ &  $5231$      \\  
  BH                     & $10^{-3}$  & $2$            &  $6\times 10^{-4}$  &  $1569$      \\  
  \hline
\end{tabular}    
\caption{The inner radii of the stellar cusp for different stellar populations $T$ calculated according to Eq.~\ref{eq_inner_radius}.} 
\label{tab_inner_radius} 
\end{table}

\begin{figure}
  \centering
  \includegraphics[width=0.5\textwidth]{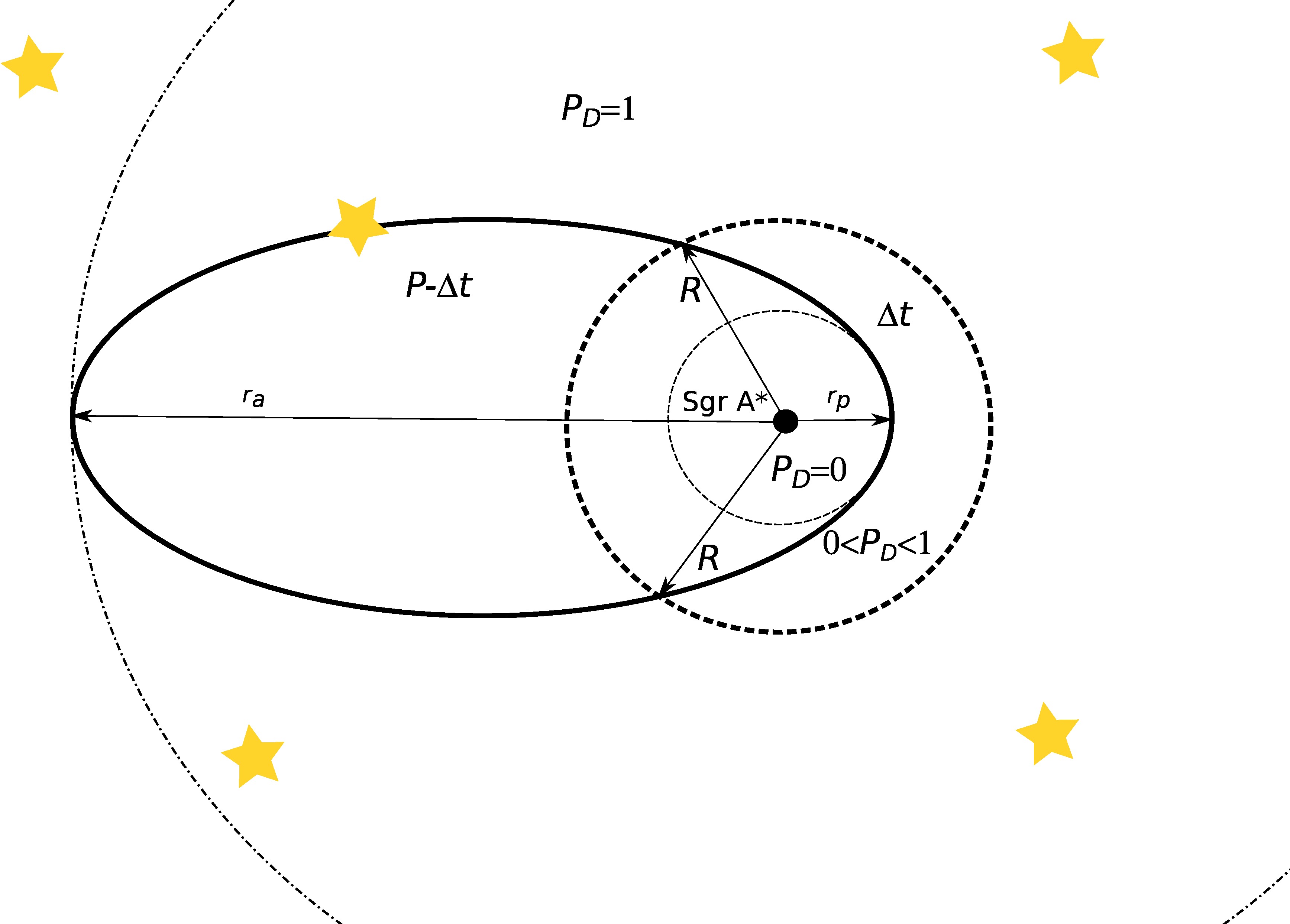}
  \caption{Basic geometry of a stellar fly-by close to Sgr~A*. The ratio of the time interval $\Delta t$ which expresses the time the star spends inside the radius $R$ and its orbital period $P$ gives the probability of detecting a star in sparse regions close to the Galactic centre.}
  \label{img_star}
\end{figure}

In these sparse regions of the Galactic centre which have a general radius $R$ with respect to the Galactic centre black hole, we may always detect with a certain probability a star whose orbital elements meet specific criteria given by the radius $R$. These criteria may be expressed in terms of the orbital elements of the star, see Fig.~\ref{img_star}. 

In order to detect a star inside the sphere of radius $R$, the orbit must be intersected by a sphere at two points with the distance $r=R$ from the focus -- black hole. In general, the distance of the star from the black hole along the elliptical orbit is given by $r=a(1-e\cos{E})$, where $a$ is the semi-major axis, $e$ is the eccentricity and $E$ is the eccentric anomaly. Since $|\cos{E}|<1$, we obtain general conditions on the orbital elements of the star $(a,e)$ and the probability of its detection inside the regions of a given radius $R$. 

The probability of detection $P_{\rm{D}}$ is non-zero, i.e. $0<P_{\rm{D}}<1$, if $r_{\rm{p}}=a(1-e)<R$ and simultaneously $r_{\rm{a}}=a(1+e)>R$, where $r_{\rm{p}}$ and $r_{\rm{a}}$ are apparently the pericentre and apocentre distances of the star from the black hole. If $r_{\rm{p}}$ approaches $R$, the probability of detection goes to zero, since the star never intersects the region of radius $R$ except for one point. On the other hand, if $r_{\rm{a}}$ approaches $R$, the probability of detection goes to one since the star is always located inside the region of radius $R$. The conditions are summarized in Table~\ref{Tabel_cond}.

\begin{table}
\centering
  \begin{tabular}{c|c|c}
    \hline
    \hline
    $P_{\rm{D}}\rightarrow 0$ & $0<P_{\rm{D}}<1$ & $P_{\rm{D}}\rightarrow 1$\\    
    \hline 
    $a(1-e)\rightarrow R$ & $a(1-e)<R<a(1+e)$ & $a(1+e)\rightarrow R$\\
    \hline
  \end{tabular}
  \caption{General conditions for the detection probability $P_{\rm{D}}$ of a star with the semi-major axis $a$ and the eccentricity $e$ inside the sphere of a given radius $R$.}
  \label{Tabel_cond}
\end{table}  

A non-trivial case is when a star passes through a region of radius $R$ for a certain time $\Delta t$. Under the assumption that we have an ideal detector (with infinite sensitivity), the probability of detecting a star at any point is given by the ratio of the time $\Delta t$, during which the star is inside $R$, and the orbital period $P_{\rm orb}$ of the star,

\begin{equation}
  P_{\rm{D}}=\frac{\Delta t}{P_{\rm orb}}\,.
  \label{eq_prob}
\end{equation}
The interval $\Delta t$ is equal to twice the time when the star is located at the distance $R$ after the peribothron passage at time $T_0$, $\Delta t=2(t(R)-T_0)$. A useful formula for the detection probability is then obtained using the Kepler equation, $M=2\pi(t(R)-T_0)/P_{\rm orb}=E(R)-e\sin{E(R)}$, where $M$ is the mean anomaly at the distance of $R$. Finally, using Eq.~\ref{eq_prob} we get,

\begin{equation}
  P_{\rm{D}}=\frac{\Delta t}{P_{\rm orb}}=\frac{1}{\pi}[E(R)-e\sin{E(R)}]\,
  \label{eq_detection_probability}
\end{equation}
where the eccentric anomaly $E(R)$ at distance $R$ can be obtained in a straightforward way from $E(R)=\arccos{\left[\frac{1}{e}\left(1-\frac{R}{a}\right)\right]}$.

For practical purposes, the eccentric anomaly expressed at the distance $R$ (when the star crosses the sphere of radius $R$), may be expressed as,

\begin{equation}
  \cos{E}(R,e,\Upsilon)=\frac{1}{e}\left[1-\left(\frac{R}{r_{\rm s}} \right)\Upsilon^{-1}(1-e) \right]\,,
  \label{eq_eccentric_anomaly1}
\end{equation}
where we expressed the radius of the field of view in Schwarzschild radii and introduced the relativistic parameter $\Upsilon=r_{\rm p}/r_{\rm s}$ \citep{2017ApJ...845...22P}, which basically represents the term, on which the post-Newtonian corrections depend. In particular, the smaller the parameter $\Upsilon$ is, the larger the periastron shift is. Using the eccentric anomaly as expressed in Eq.~\eqref{eq_eccentric_anomaly1}, the detection probability depends on three parameters, $P_{\rm D}=P_{\rm D}(R,e,\Upsilon)$. 

Even more concise representation can be obtained by introducing the parameter $\Lambda=r_{\rm p}/R$ as the ratio of the pericentre distance of a star to the field-of-view radius $R$. Then the eccentric anomaly is,

\begin{equation}
  \cos{E}(\Lambda, e)=\frac{1}{e}\left[1-\Lambda^{-1}(1-e) \right]\,,
  \label{eq_eccentric_anomaly2}
\end{equation}
which leads to the overall dependency of the detection probability $P_{\rm D}=P_{\rm D}(\Lambda,e)$. In Fig.~\ref{fig_detection_probability}, we plot the detection probability as a function of the eccentricity in the range $e=[10^{-3},0.999]$ and the parameter $\Lambda=[10^{-3},1]$. For $\Lambda>1$, the star does not enter the region, hence the detection probability is zero. Another limiting line in the parameter space $(e,\Lambda)$ is $r_{\rm a}=R$, below which the detection probability is not properly defined by Eq.~\eqref{eq_detection_probability}, but is identically equal to one, $P_{\rm D}=1$ since the whole stellar orbit for these orbital constraints lies inside the region of radius $R$. 

\begin{figure}[tbh]
  \centering
  \includegraphics[width=0.5\textwidth]{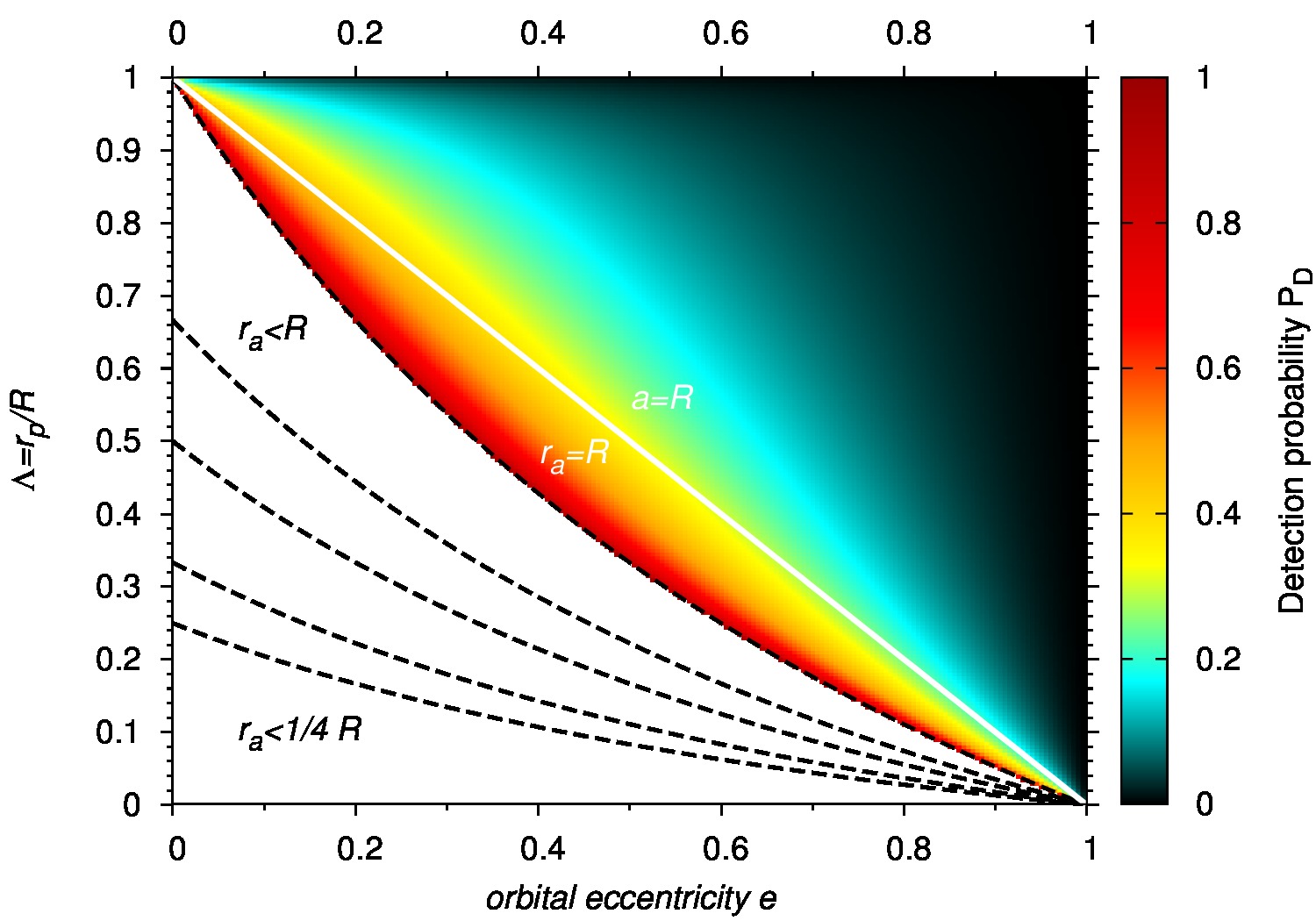}
  \caption{A colour-coded plot of the detection probability $P_{\rm D}$ of a star inside the sphere of radius $R$ (which can be understood as a circular field of view) as a function of both the orbital eccentricity $e$ and the parameter $\Lambda$, which represents the ratio of the pericentre distance $r_{\rm p}$ to $R$.}
  \label{fig_detection_probability}
\end{figure}

\begin{figure}
  \centering
  \includegraphics[width=0.5\textwidth]{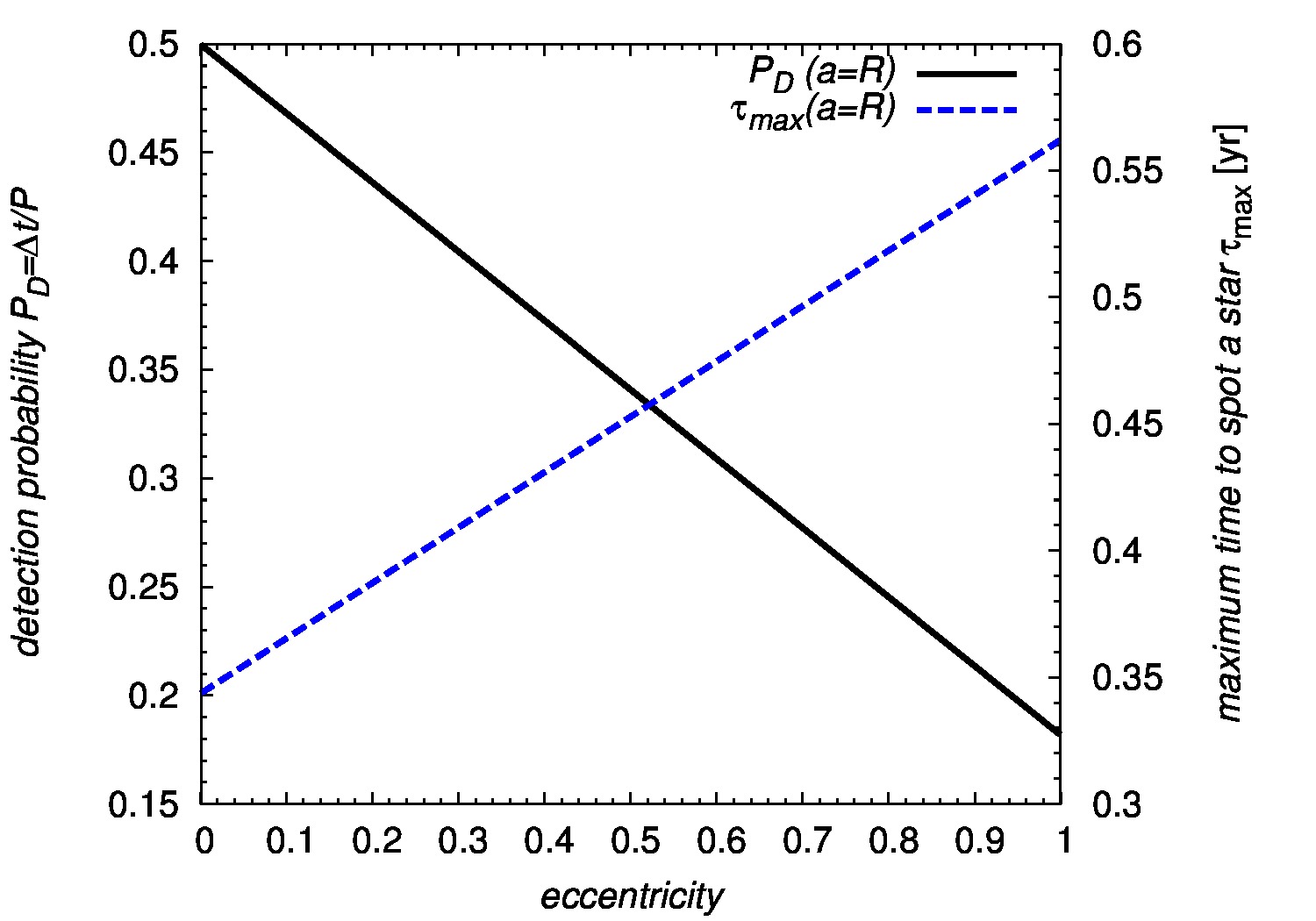}
  \caption{The relation between the detection probability (left y-axis) and the orbital eccentricity for the case $R/a=1$. The right y-axis depicts the maximum time to a spot a star $\tau_{\rm{max}}=P_{\rm{orb}}(1-P_{\rm{D}})$ as function of the eccentricity. The field-of-view is taken to be equal to the pericentre distance of S2 star $R=0.6\,\rm{mpc}=0.024\,\rm{mas}$, and the semi-major axis of a star is equal to $R$, whereas its pericentre distance is $r_{\rm{p}}=a(1-e)$.}
  \label{fig_detection_prob}
\end{figure}

The procedure for the detection probability estimate will be illustrated for a case when for a given field of view with the length-scale of $R$, we would like to know the probability of detecting a star with semi-major axis that is comparable to $R$, $a\approx R$, and the orbital period near the black hole then is $P_{\rm orb}=(4\pi^2 R^3/GM_{\bullet})^{1/2}$. This particular case is represented in Fig.~\ref{fig_detection_probability} by a solid white line. For the eccentric anomaly we get $E=\arccos{0}=\pi/2$, which then leads to the very simple relation for the detection probability $P_{\rm{D}}=1/2-e/\pi$ that depends linearly on the eccentricity $e$. The dependence of $P_{\rm D}$ on the orbital eccentricity for the case $R/a=1$ is plotted in Fig.~\ref{fig_detection_prob}, with the values along the left axis.

We also define and calculate an observationally interesting quantity $\tau_{\rm{max}}$ -- maximum time to spot a star with $a=R$ or a maximum ``waiting time'' if the star is not spotted within the field of view of radius $R$ with a given near-infrared instrument. It is simply given by $\tau_{\rm{max}}=P_{\rm orb}-\Delta t=P_{\rm orb}(1-P_{\rm{D}})$.

As an instructive case, we consider the field of view that is equal to the pericentra distance of S2 star: $R=r_{\rm{p}}^{\rm S2} \approx 0.6\,\rm{mpc}=0.024\,\rm{mas}$. This field of view is chosen not quite arbitrarily since it corresponds to the radial scale with respect to the SMBH, on which relativistic effects are important. In addition, we can show that it is expected to be practically devoid of bright stars. From Eq.~\eqref{eq_numberstars_r}, we get the total number of main-sequence stars within the periapse of S2 of the order of $N_{\star}(<r_{\rm p}^{\rm S2}) = 3.4\times 10^6 \left(r/r_{\rm h} \right)^{1.6}=9.8$. However, to obtain a number of stars bright enough to be detected, we have to multiply the total number by a fraction that follows from an IMF, $\mathrm{d}N \propto m^{-\alpha}\mathrm{d}m$, or in an integrated form,

\begin{equation}
  N_{\rm MS,det}=N_{\rm MS} \frac{m_{2}^{1-\alpha}-m_{1}^{1-\alpha}}{m_{\rm max}^{1-\alpha}-m_{\rm min}^{1-\alpha}}\,,
  \label{eq_detected_stars}
\end{equation}
where $\alpha$ is the slope of the IMF, which we consider standard \citep[$\alpha=2.35$][]{2001MNRAS.322..231K}, $(m_{\rm min}, m_{\rm max})=(0.1,100)\,M_{\odot}$ are the minimum and the maximum masses of main-sequence stars, and $(m_1,m_2)$ are the mass limits of the subset of stars of our interest. The faintest main-sequence stars in the central parsec that are detectable with current instruments have the mass of $\sim 2\,M_{\odot}$ and therefore we set $m_1=2\,M_{\odot}$ and $m_2=100\,M_{\odot}$. The number of detectable MS stars below S2 periapse falls then below one, $N_{\star, det}(<r_{\rm p}^{\rm S2}) = 3.4\times 10^6 \times 0.017 \left(r/r_{\rm h} \right)^{1.6}=0.17$. Therefore, the formula~\eqref{eq_detection_probability} for the detection of one star crossing a sparse region applies to the sphere of radius $R=r_{\rm p}^{S2}$. 

\begin{figure}[tbh]
  \centering
  \includegraphics[width=0.5\textwidth]{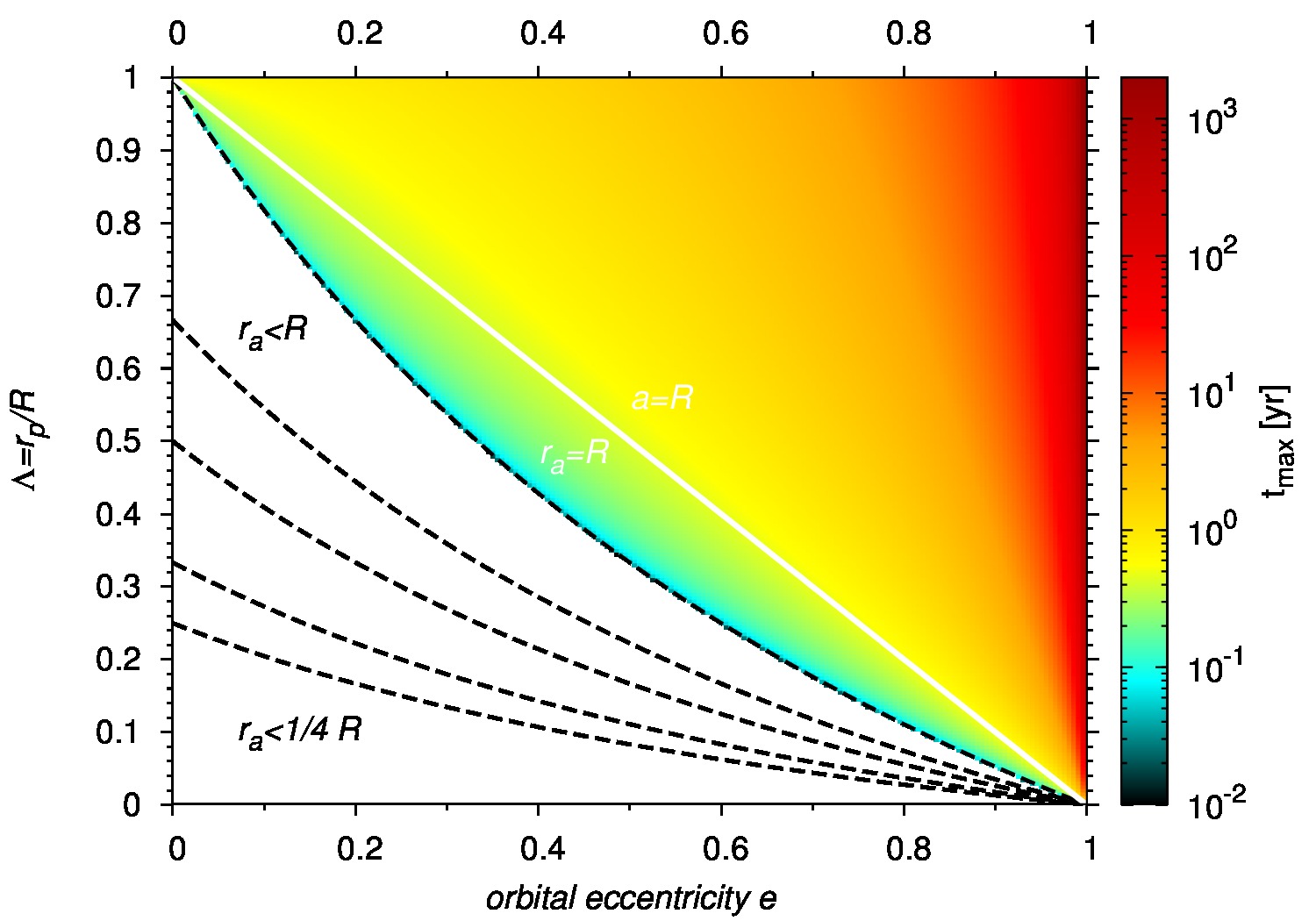}
  \caption{A colour-coded plot of the maximum time $\tau_{\rm max}$ to detect a star inside the sphere of radius $R$, which is devoid of stars, as a function of both the orbital eccentricity $e$ and the parameter $\Lambda$, which represents the ratio of the pericentre distance $r_{\rm p}$ to $R$.}
  \label{fig_detection_probability}
\end{figure}

Using Eq.~\eqref{eq_eccentric_anomaly2}, we calculate the time $\tau_{\rm max}$ as a function of the ratio $\Lambda=r_{\rm p}/R$ and the eccentricity $e$. For a specific evaluation, we consider the crossing radius of $R=r_{\rm p}^{S2}$, i.e. the pericentre distance of S2.  For a large span of ratios $\Lambda$ and eccentricities $e$, the maximum time to detect a star in a sparse region is less than or of the order of one year. Only for highly-eccentric orbits and larger ratios $\Lambda$ (when the pericentre distance is close to the radius $R$), it reaches hundreds to thousands of years. 

For the specific case $a\approx R$, we show the basic trend in Fig.~\ref{fig_detection_prob} -- the detection probability $P_{\rm D}$ decreases linearly for increasing eccentricities as expected, whereas the maximum time $\tau_{\rm{max}}$ increases in the same direction. The orbit of a star, whose semi-major axis is comparable to the pericentre distance of S2, $a\approx R \approx 1529\,r_{\rm s}$, has the orbital period of $P_{\rm orb}=(4\pi^2 R^3/GM_{\bullet})^{1/2}\simeq 0.687\,{\rm yr}$, which is short enough to spot a few orbits within several years. For very small eccentricities, the detection probability is close to $1/2$ and the maximum time to detect a star in the given field of view is close to the half of the orbital period as expected, $\tau_{\rm max} \approx 0.344\,{\rm yr}$. For increasing eccentricities, the detection probability and $\tau_{\rm max}$ behave simply linearly, as shown in Fig.~\ref{fig_detection_prob}. In case of highly-eccentric orbits with $e\approx 0.999$, the probability is $P_{\rm D}=0.182$ and the maximum ``waiting'' time is $\tau_{\rm max}=0.562\,{\rm yr}$.    

Under the assumption that within a certain volume around the Galactic centre the distribution of orbital eccentricities of stars is approximately thermalized, $n(e)\mathrm{d}e \simeq 2e\mathrm{d}e$, with which the eccentricity distribution of monitored S stars is marginally consistent \citep{2003ANS...324..535S, 2005PhR...419...65A,2010RvMP...82.3121G}, the mean eccentricity is expected to be $\overline{e} \simeq 0.67$. If we adopt this value, the pericentre distance of the star crossing the sphere with radius $R=r_{\rm p}^{\rm S2}$ is $r_{\rm p}\simeq 505\,r_{\rm s}$. Post-Newtonian effects for such a star would be easier to measure, especially the relativistic periastron advance,

\begin{equation}
  \Delta\phi(\overline{e}=0.67)=\frac{6\pi G M_{\bullet}}{c^2 a (1-e^2)}\simeq  38'\,,
  \label{eq_periastron}
\end{equation}
i.e. it would be more than half a degree, whereas for S2 star it is expected to be about one third of this value, $\Delta \phi_{\rm S2} \simeq 11.3'$.

\section{Discussion}

The derived probability $P_{\rm D}$ to detect a star in a sparse region of radius $R$ is an upper limit, i.e. we assumed that the detector has an infinite sensitivity to detect an object of a given type (a main-sequence star or a pulsar). For real detectors, in particular near-infrared telescopes or radiotelescopes, the detection threshold needs to be considered, below which the probability of detecting faint stars is naturally zero. However, for stellar flux densities above the threshold, we showed that the innermost regions close to the Galactic centre are expected to be rather sparse, with the number of bright stars, i.e. with magnitudes $<19^{m}$ in infrared $K$-band, being less than unity below the periapse distance of S2. In case no star is detected within the field of view  equal to the periapse distance of S2,$R=0.6\,{\rm mpc}$, the maximum time to spot a star is $\sim 0.3\,{\rm yr}$ for eccentricities close to zero and $\sim 0.6\,{\rm yr}$ for highly-eccentric orbits, under the assumption that the semi-major axis of a star is comparable to the radius of the field-of-view, $a\approx R$.   

A small analysis presented here shows that a number of stellar objects (bright main-sequence stars or pulsars) ideal for doing precise tests of general relativity close to Sgr~A* is rather limited. Basically, below $100$ Schwarzschild radii the total number of MS stars is $N_{\star}(<100\,r_{\rm s})\approx 0.1$ according to Eq.~\eqref{eq_numberstars}. Naturally, the expected number of bright enough stars to be detected and the number of compact remnants (pulsars) is even smaller than that. In addition, the existence of main-sequence stars is dynamically limited by the tidal disruption radius, which is $\sim 13\,r_{\rm s}$ for Solar-type stars.

Therefore, the probability to detect any stellar objects on orbits at or close to ISCO, which would be  important for distinguishing the black hole nature of Sgr~A* from other compact scenarios \citep[such as, boson stars, macroquantumness;][]{2017FoPh...47..553E}, is negligible or a matter of a coincidence. In this context, a more promising way for testing strong-gravity effects on the scale of $\sim 10\,{\rm r_{\rm s}}$ is the analysis of light curves of detected bright X-ray flares, some of which contain a substructure with the main peak and a ``shoulder'' \citep{2017MNRAS.472.4422K}. Their axisymmetric shapes can be explained to result from a flash on the length-scale of $\sim 10-20$ Schwarzschild radii. As the spot temporarily orbits around the SMBH, relativistic effects -- Doppler boosting, gravitational redshift, light focusing, and light-travel time delays-- modulate the observed signal. The mass of the central object (Sgr~A*) inferred from X-ray light curves agrees well with the mass determined from stellar orbits that are more distant by two orders of magnitude. Hence, flares and stars can complement each other on different scales. 

In this contribution, we neglected the effect of the orbital inclination, which by itself does not effect the detection probability $P_{\rm D}$ and the maximum timescale $\tau_{\rm max}$ if we consider a spherical region of radius $R$. It can, however, affect the measurement of the pericentre shift $\Delta \phi$, Eq.~\eqref{eq_periastron}, which is most difficult to be reliably measured for nearly edge-on orbits. On the other hand, the gravitational redshift $z_{\rm g}$ for the observer at infinity depends only on the distance from the black hole $r_{\rm e}$, where photons were emitted,

\begin{equation}
   z_{\rm g}(\infty, S2)=\frac{1}{\sqrt{1-r_{\rm s}/r_{\rm e}}}-1\approx 3.3\times 10^{-4}\,,
   \label{eq_gr_redshift}
\end{equation}
which is evaluated for the pericentre distance of S2, $r_{\rm e}=a_{\rm S2}(1-e_{\rm S2})\simeq 1529\,r_{\rm s}$ and leads to the radial velocity contribution to the shift of spectral lines, $v_{\rm g}=z_{\rm g}c\simeq 99\,{\rm km\,s^{-1}}$.

The other effect that was neglected were Newtonian perturbations from other stars and the stellar cluster as a whole, which, however, should be negligible in the sparse region where the number of stars is of the order of unity. In a similar way, the occurrence of blend or false stars due to the superposition of faint stars that are at the confusion limit of telescopes \citep{2012A&A...545A..70S} is expected to be rather small in the sparse region, although sources along the line of sight can still cause a certain degree of confusion. These intervening stars in the foreground can, however, be excluded based on the kinematics. 

\section{Conclusions}

We derived the probability to detect a star very close to the Galactic centre (inside the pericentre distance of S2 star), where the time-averaged number of bright stars is less than one. Considering the region of a general length-scale $R$, trivial cases for determining the detection probability $P_{\rm D}$ are when the pericentre of the stellar orbit approaches $R$, when $P_{\rm D}$ does to zero. On the other hand, the detection probability goes to one as the apocentre distance approaches $R$.

The non-trivial case is for $r_{\rm p}<R<r_{\rm a}$, when the probability is given by $P_{\rm D}=\pi^{-1}(E-e\sin{E})$, where is $E$ is the eccentric anomaly of a quasi-Keplerian stellar orbit. For stellar orbits with the semi-major axis comparable to the radius of the sparse region , $a\simeq R$, the detection probability is $P_{\rm D}=1/2-e/\pi$, i.e. it is decreasing for an increasing eccentricity. The maximum time to spot a star can then be calculated simply as $\tau_{\rm max}=P_{\rm orb}(1-P_{\rm D})$, i.e. it is larger for an increasing orbital eccentricity. For a particular and observational interesting case, when the field of view is equal to the pericentre distance of S2 star, $\tau_{\rm max}$ reaches $\sim 0.3\,{\rm yr}$ for nearly circular and $\sim 0.6\,{\rm yr}$ for highly-eccentric orbits.

To sum up, we showed that it is unlikely to detect a bright star in the innermost $R=1500$ Schwarzschild radii from Sgr~A*, where relativistic effects are prominent. However, a regular monitoring with current and future near-infrared facilities with the separation of at least $\sim 0.1\,{\rm yr}$ can yield the detection of a stellar fly-by that can be utilized as a probe of strong-gravity effects.

\section*{Acknowledgements}

We are grateful to Lorenzo Iorio for the input.
M.Z. and A.T. thank Dr. Ren\'e Hudec for organizing a high-quality conference IBWS 2018 in Karlsbad (Karlovy Vary, Czechia) based on the Bavarian-Czech partnership. M.Z. acknowledges the financial support of SFB 956 ``Conditions and impact of star formation'' at the Universities of Bonn and Cologne and MPIfR, in particular subproject A2 ``Conditions for Star Formation in Nearby AGN and QSO Hosts''. A.T. acknowledges the Czech Science Foundation Grant No. 16-03564Y and the internal grant of the Silesian University in Opava No. SGS/14/2016.

\subsection*{Author contributions}

MZ proposed the idea, made basic analysis, calculations, and all the plots and wrote the text of the manuscript. AT provided comments to the discussion of relativistic effects in the text.  

\subsection*{Conflict of interest}

The authors, MZ and AT, declare no potential conflict of interests.


\section*{Author Biography}

\begin{biography}{\includegraphics[width=60pt,height=70pt]{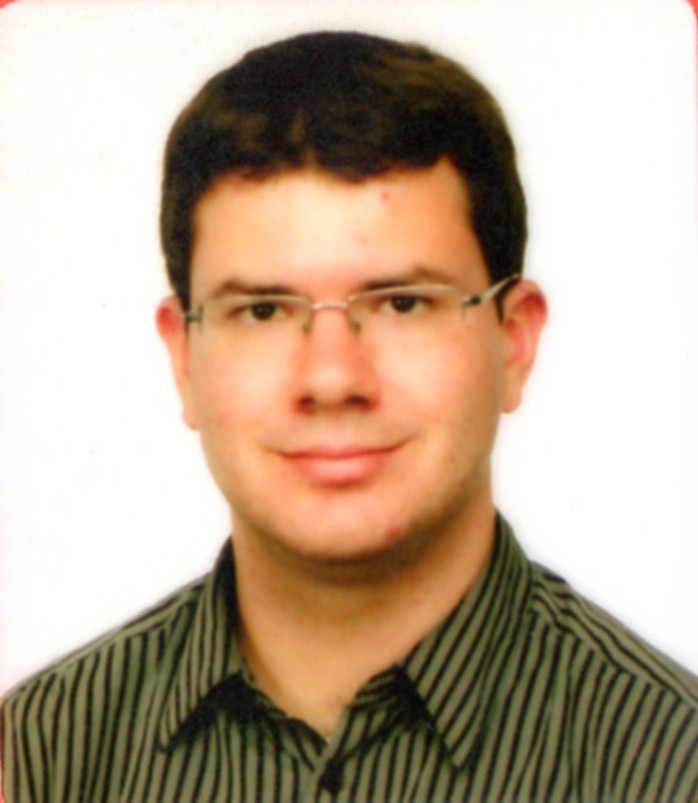}}{\textbf{Michal Zaja\v{c}ek.} Michal Zaja\v{c}ek, PhD., obtained his bachelor degree in general physics at the Charles University in Prague. He defended master thesis ``Neutron stars near the galactic centre'', supervised by Prof. Vladim\'ir Karas, at the same university in 2014. He continued with doctoral studies at the University of Cologne, Germany, and Max Planck Institute for Radioastronomy in Bonn. He defended the PhD Thesis ``Interaction between interstellar medium and black hole environment'', which was supervised by Prof. Dr. Andreas Eckart and Prof. Dr. Anton J. Zensus, in October 2017. Currently, he is a postdoctoral fellow at the Max Planck Institute for Radioastronomy in Bonn, where he focuses on studying radio-optical properties of active galactic nuclei and on the Galactic centre physics.}
\end{biography}

\end{document}